\documentclass[prd,superscriptaddress,showpacs,nofootinbib,preprintnumbers]{revtex4}

\usepackage{amsmath}
\usepackage{amsfonts}
\usepackage{graphicx}
\usepackage{dcolumn}
\def\hs{\qquad} 
\def\beq{\begin{eqnarray}} 
\def\eeq{\end{eqnarray}} 
\def\at{\left(} 
\def\aq{\left[} 
\def\cp{\right.} 
\def\ct{\right)} 
\def\cq{\right]} 
\def\lap{\Delta\,} 
\def\ii{\infty}
\def\segue{\qquad\Longrightarrow\qquad} 
\def\al{\alpha}
\def\be{\beta}
\def\ga{\gamma}
\def\de{\delta}
\def\ep{\varepsilon}
\def\ze{\zeta}

\def\si{\sigma}
\def\om{\omega}
\def\ph{\varphi}

\def\Ga{\Gamma}

\def\La{\Lambda}

\newcommand{\bea}{\begin{eqnarray}}
\newcommand{\eea}{\end{eqnarray}}
\newcommand{\beaa}{\begin{eqnarray*}}
\newcommand{\eeaa}{\end{eqnarray*}}

\def\nn{\nonumber}
\newcommand{\e}{{\rm e}}

\def\h{\tilde h}
\def\txi{\tilde\xi}
\def\B{\tilde B}
\def\C{\tilde C}
\def\R{R_0}

\begin{document}
\tolerance=5000

\title{ Vacuum energy fluctuations, the
induced cosmological constant and cosmological reconstruction in
non-minimal modified gravity models}

\author{
 Guido Cognola$\,^{(a)}$\footnote{cognola@science.unitn.it},
 Emilio Elizalde$\,^{(b)}$\footnote{elizalde@ieec.uab.es},
 Shin'ichi Nojiri$\,^{(c)}$\footnote{nojiri@phys.nagoya-u.ac.jp},
 Sergei D.~Odintsov$\,^{(b,d)}$\footnote{odintsov@ieec.uab.es also at TSPU,Tomsk}
}

\affiliation{
$^{(a)}$ Dipartimento di Fisica, Universit\`a di Trento \\
and Istituto Nazionale di Fisica Nucleare \\
Gruppo Collegato di Trento, Italia\\
\medskip
$^{(b)}$ Consejo Superior de Investigaciones Cient\'{\i}ficas
(ICE/CSIC) \, and \\ Institut d'Estudis Espacials de Catalunya
(IEEC) \\
Campus UAB, Facultat Ci\`encies, Torre C5-Par-2a pl \\ E-08193 Bellaterra
(Barcelona) Spain\\
\medskip
$^{(c)}$ Department of Physics, Nagoya University, Nagoya 464-8602, Japan\\
\medskip
$^{(d)}$ ICREA, Barcelona, Spain \, and \\
Institut de Ciencies de l`Espai (IEEC-CSIC) \\
Campus UAB, Facultat Ci\`encies, Torre C5-Par-2a pl \\ E-08193 Bellaterra
(Barcelona) Spain\\
}

\begin{abstract}
The one-loop effective action for non-minimal scalar modified gravity on
de
Sitter background with a constant scalar field is found. The corresponding
induced cosmological constant is evaluated. It is shown that quantum
effects in non-minimal modified gravity may induce an early-time de
Sitter universe even in the
situation when such solution does not occur on the classical level.
Classical reconstruction of the theory is presented in such a way that the
resulting theory has a cosmological solution unifying early-time
inflation with late-time acceleration.

 \end{abstract}


\maketitle

\section{Introduction}\label{UNO}

It has become clear recently that modified gravity may suggest a
very natural gravitational alternative for the unified description
of the early-time and the late-time
accelerating epochs of our universe (for a review and comparison of different modified
gravities, see \cite{NO-rev}). In this scenario, the universe evolution
(decrease of the universe curvature) defines the gravitational sector
which describes the evolution in its own turn.
Starting from the first modified gravity which unifies early-time
inflation with late-time acceleration, and which is consistent with the local
tests \cite{Nojiri:2003ft}, a number of realistic unified alternative
gravities of this kind has been proposed (for a list and detailed classification, see
\cite{Eli})). A very interesting sub-class of modified gravities is the
so-called
non-minimal gravity (for a review, see \cite{Nojiri:2007bt}). That is the
theory where the gravity function couples with the whole matter sector (in the
Lagrangian description). It has been proven that such non-minimal
models can easily describe the dark energy epoch \cite{matter-1}.

In the present paper we consider some cosmological and related
quantum aspects of a scalar theory with a non-minimal interaction with gravity.
More precisely, we assume the ``coupling constant'' $f(R)$
between the scalar sector and gravity to
depend explicitly on the scalar curvature. The one-loop effective action
and the corresponding induced cosmological constant for some versions of
such non-minimal gravity on the de Sitter background with a constant scalar will be
found. It is demonstrated in the following that, even when starting from flat space, the quantum
effects induce an early-time de Sitter phase (early-time inflation) in
the non-minimal gravity theory. This may be the origin of the early-time
inflation in such models. The classical cosmological reconstruction for
non-minimal theories is developed. Within such scheme, the
possibility of the unification of the inflation with the dark energy epoch
will be demonstrated.

The paper is organized as follows.
In Sect.~\ref{EFMG} we investigate the scalar non-minimal models
from the classical point of view and derive the conditions for the existence of
(anti-)de Sitter and Minkowski solutions.
The third section is devoted to the development of the classical
reconstruction scheme for the above non-minimal theory. Using such technique,
the non-minimal theory unifying early-time inflation with late-time
acceleration is reconstructed. The reconstruction of the models having the de
Sitter solutions with a time-dependent background scalar is also done.
In Sect.~\ref{OLEA}, for the scalar non-minimal model, we compute the
one-loop effective action on a maximally symmetric background with
constant scalar and derive the explicit ``on-'' and ``off-shell'' expressions.
The one-loop, gauge-fixing independent effective action is also evaluated.

Section \ref{ICC} is devoted to the calculation of the induced
cosmological constant, which can be interpreted as the vacuum energy due to
quantum fluctuations of the gravitational field and of the matter ones
on the background manifold. The one-loop effective action of the previous
section is used in this calculation.
In section six we study the induced cosmological constant for several
non-minimal models on the de Sitter background in the large-curvature limit.
It is explicitly shown there that the one-loop effective action may induce the
de Sitter space at the quantum level, even in the situation when such classical
solution is absent. Some summary is presented in the final section.

\section{ Scalar non-minimal modified gravity}\label{EFMG}

We start from the model introduced in
Ref.~\cite{matter-1,Allemandi:2005qs},
which is characterized by a non-minimal coupling between
gravity and the scalar field sector. It has been proposed as
a viable model of modified gravity \cite{matter-1}, which can describe
the late-time acceleration epoch. Different aspects of such theory have been
studied, including the induced cosmological constant \cite{DC}, some cosmological
aspects, and the appearance of an extra force \cite{criteria-1,Faraoni:2007sn} (for a
review, see \cite{Nojiri:2007bt}). The action is taken to be
\beq
\label{Act1}
S=-\int d^4 x\,\sqrt{-g}\aq-R+f(R)L_S\cq\,,
\eeq
where
\beq
\label{AA1}
L_S=-\frac12\,g^{ij}\partial_i\phi\partial_j\phi-V(\phi)
\eeq
is the Lagrangian density of a scalar field
$\phi$ with an arbitrary potential $V(\phi)$.
The classical field equations following from (\ref{Act1}) read
\beq\label{GijS}
R_{ij}-\frac12\,Rg_{ij}=\frac12\,f(R)T^S_{ij}
+\aq g_{ij}\,\Box-\nabla_i\nabla_j+R_{ij}\ct\aq f'(R)L_S\cq\,,
\eeq
\beq\label{Sc0}
-\frac{1}{\sqrt{-g}}\,\partial_i\aq\sqrt{-g}\,
 f(R)g^{ij}\partial_j\phi\cq+f(R)V'(\phi)=0\,,
\eeq
where
\beq
T^S_{ij}=
g_{ij}\aq\frac12\,g^{rs}\partial_r\phi\partial_s\phi+V(\phi)\cq
-\partial_i\phi\partial_i\phi
\eeq
is the standard energy-momentum tensor of the scalar field.

The action (\ref{Act1}) has a
(anti)de Sitter solution $(\R,\phi_0)$, with constant curvature
$\R$ and constant field $\phi_0$, if the following conditions
are fulfilled:
\beq\label{dSG}
\R-\R\,f'(\R)V(\phi_0)+2f(\R)V(\phi_0)=0\,,
 \hs\hs f(\R)V'(\phi_0)=0\,.
\eeq
In order to satisfy the last condition we can choose
$V'(\phi_0)=0$ or $f(\R)=0$. We also note that, if
$V(\phi_0)=0$, then the solution is flat, but one can
have Minkowski space solution without restrictions on the potential,
if $f(0)=0$.

The stability of the solutions above with respect to homogeneous perturbations
can be obtained by using method of
Refs.~\cite{Cognola:2007vq,Cognola:2008wy}.
One easily gets
\beq
\frac{1+f'(\R)V(\phi_0)}{\R f''(\R V(\phi_0)}>1\,.
\eeq
The
field equation for the scalar field can be written in the form
\beq
\Box\phi-V'(\phi)=0\,,
\eeq
It is interesting to note that in the case of vanishing potential there
are no solutions with constant $R=\R$,
since in such a case from Eqs.~(\ref{GijS}) and (\ref{Sc0}) we get
\beq
\label{I0}
\Box\phi=0\,,\hs\hs
 R+\aq 2f(R)-Rf'(R)\cq\,\frac{g^{ij}\partial_i\phi\partial_j\phi}2=0\,.
\eeq
In the FRW background, the first equation has the following form:
\beq
\label{Ia}
\Box \phi = \frac{1}{a^3}\frac{d}{dt}\left(a^3 \frac{d\phi}{dt}\right) = 0\,,
\eeq
which can be solved as
\beq
\label{Ib}
\frac{d\phi}{dt} = C a^{-3}\,.
\eeq
Here $C$ is a constant of integration and $a$ is a scale factor.
Then the second equation in (\ref{I0}) has the following form:
\beq
\label{EdS}
-\frac{2\R}{2f(\R)-\R f'(\R)} = g^{ij}\partial_i\phi \partial_j\phi
= - C^2 a^6\,.
\eeq
In the de Sitter background, the l.h.s. is a constant but the scale factor $a$ is
not a constant in the expanding universe. Then, unless $C\neq 0$, there is not consistent solution
but $C\neq 0$ implies $R=0$, which is the flat space-time.

The stability of the solution can be obtained by using \cite{Cognola:2008wy}.
We have
\beq\label{STdS}
\frac{1+a^2f'(\R)}{a^2\R f''(\R)}-1>0\,,
\eeq
which agrees with the standard result when $a^2=1$.

\section{Cosmological reconstruction and the unification of the inflation epoch
with dark energy}

In this section we consider how one can construct a classical model
which reproduces an almost arbitrary
time evolution of the universe. Note that a general scheme of cosmological
reconstruction of modified gravity has been developed for a number of
modified gravities (for review, see \cite{review} based on the technique
developed in Refs.~\cite{reconst1,reconst2}).
The same method will be applied below. The starting action is
(\ref{Act1}) with (\ref{AA1}).
We redefine the scalar field by using an adequate function $h$ as
$\phi=h(\varphi)$. Then Eq.~(\ref{AA1}) has the following form:
\beq
\label{AA2}
L_S=-\frac12\,\omega(\varphi) g^{ij}\partial_i\varphi\partial_j\varphi - \tilde V(\varphi)\,,
\quad \omega(\varphi) \equiv h'(\varphi)^2\,,\quad
\tilde V(\varphi) \equiv V\left(h\left(\varphi\right)\right)\,.
\eeq
Especially by choosing $h$ properly, one may identify $\varphi$ as a time
coordinate.

In the FRW background with a flat spatial part, the equation corresponding
to the first FRW equation has the following form:
\beq
\label{AA3}
3H^2 = \frac{f(R)}{2}\left( \frac{\omega(\varphi)}{2}{\dot \varphi}^2 + \tilde V(\varphi )
\right) - 3 \left\{ H\frac{d}{dt} - \left(\dot H + H^2 \right) \right\} \left\{ f'(R) \left(
\frac{\omega(\varphi)}{2}{\dot \varphi}^2 - \tilde V(\varphi) \right) \right\}\,.
\eeq
Here the scalar curvature is $R= 6 \dot H + 12 H^2$. On the other hand, by
 variation over $\varphi$, we find
\beq
\label{AA4}
0 = f(R) \left(\frac{\omega'(\varphi)}{2}{\dot \varphi}^2 - \tilde V' (\varphi) \right)
 - \left(\frac{d}{dt} + 3H \right)\left\{f(R) \omega(\varphi) \dot \varphi \right\}\,.
\eeq
Identifying $\varphi$ as a cosmological time coordinate $\varphi = t$, Eqs.~(\ref{AA3}) and
(\ref{AA4}) can be rewritten as
\bea
\label{AA5}
3H^2 &=& \frac{f\left(R\left(t\right)\right)}{2}\rho(t)
 - 3\left\{ \frac{d}{dt} - \left(\dot H + H^2 \right)\right\}
\left( f'\left(R\left(t\right)\right) p(t) \right)\,, \\
\label{AA6}
0 &=& f\left(R\left(t\right)\right) p' (t) - \left(\frac{d}{dt} + 3H\right)
\left\{ f\left(R\left(t\right)\right) \left( p(t)
+ \rho(t) \right)\right\} \,.
\eea
Here
\beq
\label{AA7}
\rho(\varphi) \equiv \frac{\omega(\varphi)}{2} + \tilde V(\varphi)\,, \quad
p(\varphi) \equiv \frac{\omega(\varphi)}{2} - \tilde V(\varphi)\,.
\eeq
In the following, we consider the cosmological reconstruction of the
model (for a general review of reconstruction in different modified
gravities, see \cite{review}), that is, given the specific time-development
of the scale factor $a$ or the Hubble rate $H$, we construct a model which
reproduces such evolution. Although only one function $a=a(t)$ or
$H=H(t)$ is given, we have
two functions $f(R)$ and $V(\phi)$ in the action (\ref{Act1}). Then,
in principle one of the functions can be arbitrary. For example, if we
choose $f(R)=1$, we have the usual scalar-tensor theory \cite{F-M}.
Now we choose
\beq
\label{AA8}
f(R) a^3 = f_0\,.
\eeq
Here $f_0$ is a constant. If one solves $R=R(t)$ with respect to $t$ as
$t=T(R)$ for a given scale
factor $a=a(t)$, $f(R)$ is explicitly given by $f(R) = f_0/ a\left(T\left(R\right)\right)^3$.
For example, the scale factor of the universe with one kind of perfect fluid, with
 constant EoS parameter $w$, is given by $a\propto t^{2/3(1+w)}$; then
$F(R) \propto R^{-1/3(1+w)}$.
Note that we cannot choose (\ref{AA8}) for the de Sitter solution with $f(R)$
being a constant, since $R$ is a constant but $a$ is not in de Sitter universe.
In the asymptotic de Sitter universe,
where $R$ is not exactly constant, we can choose (\ref{AA8}) as we will later
see.

For given $a=a(t)$ or $H=H(t)$, which yields $R=R(t)$, Eqs.~(\ref{AA5}) and (\ref{AA6}) can be
solved with respect to $\rho(\phi)$ and $p(\phi)$, as follows
\bea
\label{AA9}
\rho(\varphi) &=& \left[ \frac{6a\left(t\right)^3}{f_0} \left\{
H(t)^2 - \frac{Q(t)}{a(t)^3} \right\}\right]_{t=\varphi}\,,\\
\label{AA10}
p(\varphi) &=& - \left[ a(t) H(t) \int^t \frac{Q(t')dt'}{a(t')^4 H(t')^2} \right]_{t=\varphi}\,,\\
\label{AA11}
Q(t) &\equiv & \int^{t} dt' a(t')^3 \left( 2H(t') \dot H(t') + 3 H(t')^3 \right)\,.
\eea
 From $\rho(\varphi)$ and $p(\varphi)$, one finds the explicit form of
$\omega(\varphi)$ and $\tilde V(\varphi)$, by using (\ref{AA7}).

As an example, we consider the following Hubble rate
\beq
\label{AA12}
H (t) = \frac{H_L \e^{\alpha t} + \beta H_I}{\e^{\alpha t} + \beta}\,.
\eeq
Here, $H_L$, $H_I$, $\alpha$, and $\beta$ are positive constants.
When $t\to -\infty$, we find $H(t) \to H_I$, which may be identified with the inflation of
the early universe and, when $t\to +\infty$, $H(t)\to H_L$, which may be
identified with the accelerated expansion of the late universe.
Eq.~(\ref{AA12}) yields
\bea
\label{AA13}
a(t) &=& a_0 \e^{H_I t} \left(\e^{\alpha t} + \beta \right)^{\left(H_L - H_I\right)/\alpha}\,, \nn \\
R(t) &=& \frac{12 H_L^2 \e^{2 \alpha t}
+ \left\{24 \beta H_L H_I + 6\alpha \beta \left(H_L - H_I\right)\right\} \e^{\alpha t}
+ 12 \beta^2 H_I^2}{\left(\e^{\alpha t} + \beta\right)^2}\,,
\eea
which give the forms of $f(R)$, $\omega(\varphi)$, and
$\tilde V(\varphi)$ by using (\ref{AA7}), (\ref{AA8}), (\ref{AA9}), and (\ref{AA10}).

We now consider the asymptotic form of $f(R)$, $\omega(\varphi)$, and $V(\varphi)$, in the limit
$t\to \pm\infty$, for the model in (\ref{AA12}). For simplicity, we may choose $\alpha = H_1$
and $\beta = 1$.
First, we consider the behaviors of $f(R)$, $\omega(\varphi)$, and $V(\varphi)$ when $t\to -\infty$.
Since $H_I \gg H_L$, we find
\beq
\label{AAA1}
H(t) \sim H_I \left( 1 - \e^{H_I t} \right)\,,\quad
a(t) \sim a_0 \e^{H_I t} \left( 1 - \e^{H_I t} \right)\,,\quad
R \sim 12 H_I^2 \left( 1 - \frac{5}{2} \e^{H_I t} \right)\,,
\eeq
which give
\bea
\label{AAA2}
&& f(R) \sim \frac{f_0}{a_0^3} \left\{ \frac{2}{5} \left( 1 - \frac{R}{12 H_I^2} \right) \right\}^{-3}
\left\{ 1 + \frac{6}{5} \left( 1 - \frac{R}{12 H_I^2} \right) \right\}\,, \nn \\
&& \rho(\varphi) \sim \mathcal{O} \left(\e^{5H_I t} \right)\,, \quad
p \sim 1 - H_I \left( \varphi - \varphi_0 \right) \e^{H_I \varphi} \,.
\eea
Here $\varphi_0$ is a constant of integration.
Then, one gets
\beq
\label{AAA3}
\omega(\varphi) \sim - \frac{\tilde V (\varphi)}{2} \sim
1 - H_I \left( \varphi - \varphi_0 \right) \e^{H_I \varphi} \,.
\eeq
Note that $R$ could be large but always $R<12 H_I^2$.

On the other hand, when $t\to +\infty$ so that $\e^{-H_Lt} \ll \frac{H_L}{H_I} \ll 1$, we find
\beq
\label{AAAA1}
H(t) \sim H_L + H_I\e^{-H_I t}\,,\quad
a(t) \sim a_0 \e^{H_L t}\left(1 - \e^{-H_I t} \right)\,,\quad
R \sim 12 H_L^2 - 6 H_I^2 \e^{-H_I t}\,,
\eeq
which give
\bea
\label{AAAA2}
&& f(R) \sim \frac{f_0}{a_0^3} \left( \frac{12H_L^2 - R}{6H_I^2} \right)^3
\left( 1 - \frac{R}{3H_I^2} \right)\,, \nn \\
&& \rho(\varphi) \sim \mathcal{O} \left(\e^{\left(3H_L - H_I\right) t} \right)\,, \quad
p \sim \frac{H_L}{H_I} + \e^{-H_I t}\,.
\eea
Then, one obtains
\beq
\label{AAAA3}
\omega(\varphi) \sim - \frac{\tilde V (\varphi)}{2} \sim \frac{H_L}{H_I} + \e^{-H_I t} \,.
\eeq
Thus, the function interpolating between the two functions $f$ above describes
inflation, when $t\to -\infty$, and the late-time acceleration epoch, when $t\to
+\infty$. Its explicit form is however quite complicated. But in any case, such theory describes
a perfectly unified evolution from the early-time inflation era to the late-time
acceleration one.

Instead, when we choose the condition (\ref{AA8}), if we consider a general $f(R)$,
by deleting $\rho(t)$ in (\ref{AA5}) and (\ref{AA6}), we obtain
\beq
\label{AA14}
0= p(t) \left( - 3H f\left(R\left(t\right)\right) - \frac{df\left(R\left(t\right)\right)}{dt} \right)
 - \frac{6}{a(t)^3}\frac{d}{dt}\left(a^4 H^2 \frac{d}{dt}\left(\frac{f'\left(R\left(t\right)\right)
p(t)}{a(t)H(t)}\right)\right) - \left( 12 H \dot H + 18 H^3 \right) \,.
\eeq
If $f(R)$ is properly given, Eq.~(\ref{AA14}) can be regarded as a
differential equation for $p(t)$ (compare with \cite{review,reconst1}.
If we find the form of $p(\varphi)$, by solving (\ref{AA14}), we also find
the form of $\rho(\varphi)$ by using (\ref{AA7}), as
\beq
\label{AA15}
\rho(\varphi) = \left. \frac{6}{f\left(R\left(t\right)\right)} \left[H(t)^2
+ \left(H(t) \frac{d}{dt} - \dot H(t) - H(t)^2 \right)\left\{f'\left(R\left(t\right)\right) p(t)\right\}
\right]\right|_{t=\varphi}\,.
\eeq
For an example, we consider the de Sitter space where $H$ is a constant $H=H_0$.
Then, Eq.~(\ref{AA14}) reduces to
\beq
\label{AA16}
0= \frac{d^2 p(t)}{dt^2} + 2H_0 \frac{dp(t)}{dt}
+ \left( - 3H_0^2 + \frac{f\left(12H_0^2\right)}{2f'\left(12H_0^2\right)}\right)p(t)
+ \frac{3H_0^2}{f'\left(12H_0^2\right)}\,,
\eeq
which can be solved as
\bea
\label{AA17}
p(t) &=& C_+ \e^{\lambda_+ t} + C_- \e^{\lambda_- t}
+ \frac{3H_0^2}{6H_0^2 f'\left(12H_0^2\right) + f\left(12H_0^2\right)}\,, \nn \\
\lambda_\pm &\equiv & H_0 \pm \sqrt{4 H_0^2
 - \frac{f\left(12H_0^2\right)}{2f'\left(12H_0^2\right)}} \,.
\eea
Here, the $C_\pm$ are constants of integration.
Then, Eq.~(\ref{AA15}) gives
\beq
\label{AA18}
\rho(t) = \frac{6H_0^2}{f\left(12H_0^2\right)}\left( \frac{3 H_0^2 f'\left(12H_0^2\right)
+ f \left(12H_0^2\right)}
{6 H_0^2 f'\left(12H_0^2\right) + f \left(12H_0^2\right)}
+ \left(H_0 \lambda_+ - H_0^2\right) C_+ \e^{\lambda_+ t}
+ \left(H_0 \lambda_- - H_0^2\right) C_+ \e^{\lambda_- t} \right)\,.
\eeq
In particular, in the case when the constants of integration vanish, $C_\om = 0$, both $p(t)$
and $\rho$ become constant, and then $\omega$ and $V$ are given by
\bea
\label{AA19}
&& \omega = \rho + p
= \frac{9H_0^2\left( 2 H_0^2 f'\left(12H_0^2\right)
+ f \left(12H_0^2\right)\right)}
{f\left(12H_0^2\right)\left( 6 H_0^2 f'\left(12H_0^2\right) + f \left(12H_0^2\right)\right)} \,,\\
&& V = \frac{\rho - p}{2}
= \frac{3H_0^2\left( 6 H_0^2 f'\left(12H_0^2\right)
+ f \left(12H_0^2\right)\right)}
{f\left(12H_0^2\right)\left( 6 H_0^2 f'\left(12H_0^2\right) + f \left(12H_0^2\right)\right)}
\,.
\eea
We should not fix $\varphi$, and therefore $\phi$ is not a constant
\beq
\label{AA20}
\varphi = t\,,\quad \phi = \sqrt{\omega} t + \phi_0\,.
\eeq
Then the above de Sitter solution does not correspond to any of the four classes
of dS solutions (\ref{dSC1-4}) for which the one-loop effective action is
evaluated in the next section.

\section{One-loop quantization around a maximally symmetric space}\label{OLEA}

We will here perform the one-loop quantization
of the (Euclidean) classical scalar model we have discussed above.
One-loop contributions are certainly important, especially during
the inflationary phase, but as it has been
shown in \cite{Cognola:2005sg},
they also provide a very powerful method in order to study
the stability of the solutions.

In accordance with the background field method, we consider
small fluctuations $(h_{ij},\ph)$ of the fields $(g_{ij},\phi)$,
around a maximally symmetric space. Then,
\beq
g_{ij}\to g_{ij}+h_{ij}\,,\hs\phi\to\phi+\ph\,,\hs
R_{ijrs}\to R_{ijrs}+O(\nabla h_{ij})\,.
\eeq
Here and in what follows $g_{ij}$ is the metric
of $SO(4)$ and $R_{ijrs}$ the corresponding Riemann tensor.
Thus
\beq
R_{ijrs}=\frac{\La}{3}\at g_{ir}g_{js}-g_{is}g_{jr}\ct\,,
 \hs R_{ij}=\La\,g_{ij}\,,
 \hs R=4\La = \mbox{const}.
\eeq
As already said above, if the functions $f(R)$ and $V(\phi)$ satisfy the conditions
(\ref{dSG}), then $(g_{ij},\phi)$ is a classical solution
of (\ref{Act1}) with constant $\phi=\phi_0$ and $R=\R$.
We can distinguish the following four cases:
\beq
\begin{array}{lll}
 a)\hs f(\R)=0\,,&\hs f'(\R)V(\phi_0)=1\,,&\hs\R\neq0\,, \\
 b)\hs V'(\phi_0)=0\,,&\hs \R-\R f'(\R)V(\phi_0)+2f(\R)V(\phi_0)=0\,,&\hs\R\neq0\,, \\
 c)\hs f(0)=0\,,& &\hs\R=0\,, \\
 d)\hs V'(\phi_0)=0\,,&\hs V(\phi_0)=0\,,& \hs\R=0\,. \\
\end{array}
\label{dSC1-4}\eeq
The last two cases correspond to Minkowski solutions.

Now, we perform a Taylor expansion of the action around
the $SO(4)$ background manifold, up to second order in the small
variables $(h_{ij},\ph)$.
A straightforward calculation \cite{Cognola:2005de,Cognola:2006sp,Cognola:2009jx}
for zero and second order contributions in the fields $(h_{ij},\ph)$ and
disregarding total derivatives yields, respectively,
\beq\label{E0}
{\cal L}_0&=&-R-fV\segue I
=\int_{SO(4)}dx^4\,\sqrt{g}\,{\cal L}_0
=-\frac{24\pi^2[R+V\,f(R)]}{\La^2}
\eeq
and
\beq\label{Ea}
{\cal L}_2&=&
 -\frac12\,f\,\ph\,(-\lap+V'')\,\ph
 -V'\,\ph\,\aq\frac{3f'}4\,\at-\lap-\frac{R}3\ct
 +\frac12\,f\cq\,h
 +\frac{3f'}4\,V'\,\ph\at-\lap-\frac{R}3\ct\lap\si
\nn\\&&\hs
 -\frac{3}{32}\,h\aq 3Vf''\,\at-\lap-\frac{R}3\ct^2
 +Vf'\,\at-\lap-\frac{2R}3\ct
 +\at-\lap+\frac{2Vf}3\ct\cq\,h
\nn\\&&\hs
 +\frac{3f'V'}{16}\,h\at-\lap-\frac{R}3\ct\,
 \aq3f''V\at-\lap-\frac{R}3\ct+f'V+1\ct\lap\si
\nn\\&&\hs
 -\frac3{32}\,\si\at-\lap-\frac{R}3\ct\,\aq
 3f''V\at-\lap-\frac{R}3\ct\,\lap-f'V(-\lap+R)
 +\lap+R+2fV\cq\lap\si
\nn\\&&\hs
 +\frac{R+2fV-f'R V}4\,
 \txi^k\,\at-\lap-\frac{R}4\ct\txi_k
\nn\\&&\hs
+\frac14\,\h^{ij}\aq-\lap+\frac{2R}3+fV
 +f'V\at-\lap-\frac{R}3\ct\cq\,\h_{ij}\,,
\eeq
$I$ being the classical action on $SO(4)$.
In contrast with Sect.~(\ref{EFMG}),
here $\nabla_k$ and $\Delta=g^{ij}\nabla_i\nabla_j$
represent the covariant derivative
and the D'Alembertian in the unperturbed metric $g_{ij}$,
and all functions are evaluated on the background
$(g_{ij},\phi)$. This means that $f=f(R)$,
$f'=f'(R)$, and so on.

For technical reasons we have also carried out
the standard expansion of the tensor field $h_{ij}$ in
irreducible components \cite{frad}, that is
\beq h_{ij}&=&\h_{ij}+\nabla_i\txi_j+\nabla_j\txi_i
 +\nabla_i\nabla_j\sigma+\frac14\,g_{ij}(h-\lap\sigma)\:,
\label{tt}\eeq
where $\si$ is the scalar component, while $\txi_i$
and $\h_{ij}$ are the vector and tensor components, respectively.
They have the properties
\beq
\nabla_i\txi^i=0\:,\hs \nabla_i\h^i_j=0\:,\hs
\h^i_i=0\:. \label{AAA4}
\eeq

Invariance under diffeomorphisms
renders the operator in the $(h,\si)$ sector non-invertible.
One needs to involve a gauge fixing term and a corresponding
ghost compensating term.
We will consider the class of gauge conditions:
\beq
\chi_k=\nabla_j h_{jk}-\frac{1+\rho}4\,\nabla_k\,h\:,
\nn
\eeq
parameterized by the real parameter $\rho$ and,
for gauge fixing, we choose the quite general form \cite{buch}
\beq
{\cal L}_{gf}=\frac12\,\chi^i\,G_{ij}\,\chi^j\,,\hs\hs
G_{ij}=\ga\,g_{ij}+\beta\,g_{ij}\lap\,,
\label{AAA5}
\eeq
where the term proportional to $\ga$ is the one normally
used in Einstein's gravity.

Finally, one also has to add the ghost Lagrangian \cite{buch}
\beq
{\cal L}_{gh}= B^i\,G_{ik}\frac{\de\,\chi^k}{\de\,\ep^j}C^j\,,
\label{AAA6}
\eeq
where $C_k$ and $B_k$ are the ghost and anti-ghost
vector fields, respectively, while $\de\,\chi^k$
is the variation of the gauge condition due to
an infinitesimal gauge transformation of
the field (details can be found
in Refs.~\cite{buch,Cognola:2005de}).

In order to compute the one-loop contributions to
the effective action, one has
to consider the path integral for the
bilinear part
${\cal L}= {\cal L}_2+\,{\cal L}_{gf}+{\cal L}_{gh}$
of the total Lagrangian and take into
account the Jacobian due to the change
of variables with respect to
the original ones.
In this way, we get \cite{frad,buch}
\beq
\label{Z1}
Z^{(1)}&=&\at\det G_{ij}\ct^{-1/2}\,\int\,D[h_{ij}]D[C_k]D[B^k]\:
\exp\,\at\int\,d^4x\,\sqrt{g}\,{\cal L}\ct
\nn\\
&=&\at\det G_{ij}\ct^{-1/2}\,\det J_1^{-1}\,\det J_2^{1/2}\,
\nn\\
&&\times \int\,D[h]D[\h_{ij}]D[\txi^j]D[\si] D[\C_k]D[\B^k]D[c]D[b]
\:\exp\, \at\int\,d^4x\,\sqrt{g}\,{\cal L}\ct\,,
\eeq
where $J_1$ and $J_2$ are the Jacobians coming from the
change of variables in the ghost and tensor sectors, respectively.
They read \cite{buch}
\beq
J_1=\lap_0\,,\hs\hs
J_2=\at-\lap_1-\frac{R}{4}
\ct\at-\lap_0-\frac{R}{3}\ct\,\lap_0\,,
\label{AAA13}
\eeq
while the determinant of the operator $G_{ij}$
acting on vectors in our gauge assumes the form (up to a constant)
\beq
\det G_{ij}=
 \det\at\lap_1+\frac{\ga}{\beta}\ct\,
 \det\at\lap_0+\frac{R}4+\frac{\ga}{\beta}\ct\,,
\label{AAA14}
\eeq
and it is trivial in the simplest case, $\beta=0$.
By $\lap_n$ we indicate the Laplacians acting on
transverse tensors fields of order $n$.

A straightforward computation, disregarding zero modes of gravity and
the multiplicative anomaly \cite{Elizalde:1997nd},
leads to the one-loop contribution
$Z^{(1)}(\ga,\be,\rho)$ to the ``partition function''.
The result is a very complicated expression depending on
the gauge parameters. As is well known, on shell---that is, when
one of the conditions in (\ref{dSC1-4}) is satisfied---$Z^{(1)}$
does not depend on the gauge.
For the four possible cases we get, respectively,
\beq
\label{PFos1-4}
\begin{array}{l}
a)\hs Z^{(1)}_{\rm on-shell}=\aq\frac{\det\at-\lap_1-\La_0\ct}
 {\det\at\,-\lap_2+\frac23\,\La_0\ct}\cq^{1/2}\,
 \aq\det\at\,-\lap_0-\frac43\,\La_0\ct\cq^{-1/2}\,, \\
b)\hs Z^{(1)}_{\rm on-shell}=\aq\frac{\det\at-\lap_1-\La_0\ct}
 {\det\at\,-\lap_2+\frac23\,\La_0\ct}\cq^{1/2}\,
 \aq\det\at\,-\lap_0+V''_0\ct\,
 \det\at\,-\lap_0+M_0^2\ct\cq^{-1/2}\,, \\
c)\hs Z^{(1)}_{\rm on-shell}=\aq\frac{\det\at-\lap_1\ct}
 {\det\at\,-\lap_2\ct}\cq^{1/2}\,, \\
d)\hs Z^{(1)}_{\rm on-shell}=\aq\frac{\det\at-\lap_1\ct}
 {\det\at\,-\lap_2\ct}\cq^{1/2}\,
 \aq\det\at\,-\lap_0+V''_0\ct\cq^{-1/2}\,.
\end{array}
\eeq
Here we have introduced an effective mass
\beq
\label{M0}
M_0^2=\frac{8f''_0\La_0^2V_0-f_0V_0-4\La_0}
 {6f''_0\La_0 V_0}\,,
\eeq
$\La=\La_0=\R/4$, $\R$ being the solution of
(\ref{dSC1-4}), and $f_0=f(\R)$, $V_0=V(\phi_0)$, and so on.
We see that, in the second case ($b$),
the scalar potential explicitly appears in the
expression of the on-shell one-loop effective action.
It looks like, in the first case ($a$),
the scalar field does not give a contribution, but
this is not so, since in pure
Einsteinian gravity with a cosmological constant the last
term in equation ($a$) is not present. This is a contribution
due to the non-minimal coupling between gravity and matter.

 From expressions in (\ref{PFos1-4}) one can obtain the stability condition
of the corresponding de Sitter solution, by requiring that all Laplacian-like
operators involved have no negative eigenvalues \cite{Cognola:2005sg}.
For example, the stability of de Sitter solution corresponding to the
second case ($b$) is assured if $V''_0>0$ and $M_0^2>0$, which are the conditions one obtains
by the classical expression (\ref{STdS}). On the contrary, the solution
corresponding to the first case ($a$) is always unstable, in contrast with
what one obtains from (\ref{STdS}). But it has to be stressed that the classical condition
(\ref{STdS}) has been derived by considering only homogeneous perturbations.

As already explained above, the off-shell one-loop contribution
to the partition function, in the most general case, is a very complicated
expression we shall not write here explicitly. For the sake of simplicity
we will restrict the gauge parameters by putting $\be=0$
and we shall consider only the two special cases $V(\phi)=V'(\phi)=0$, $V''(\phi)\neq0$
and, alternatively, $f(R)=0$, but $f'(R)\neq0$, $f''(R)\neq0$.
In this way, the expression of
$Z^{(1)}(\ga,0,\rho)$ notably simplifies. We get
\beq
Z^{(1)}(\ga,0,\rho)&=&
 \det\at-\lap_0-\frac{4\La}{3-\rho}\ct\,
 \det\at-\lap_1-\La\ct\,
\nn\\&&\hs\times
 \aq\det\at-\lap_2+\frac83\La\ct\,
 \det\at-\lap_1-\La\aq1-\frac2\ga\cq\ct\,
 \cp
\nn\\&&\hs\hs\times
 \aq\det\at-\lap_0-\frac{4\La[2(\ga-1)+\ga(1-\rho)^2]}
 {\ga(3-\rho)^2}\ct\,
 \det\at\,-\lap_0+V''\ct\,\cq^{-1/2}\,,
\label{PF1}\eeq
\beq
Z^{(1)}(\ga,0,\rho)&=&
 \det\at-\lap_0-\frac{4\La}{3-\rho}\ct\,
 \det\at-\lap_1-\La\ct\,
\nn\\&&\hs\times
 \aq\det\at-\lap_2-\frac{4\La(f'V-2)}{3(f'V+1)}\ct\,
 \det\at-\lap_1-\La+\frac{2\La(f'V-1)}{\ga}\ct\,
 \cq^{-1/2}\,
\nn\\&&\hs\hs\times
 \aq\det\at-\lap_0-\frac{4\La}3\ct\,
 \det\at-\lap_0-\frac{\La}3\,q_1\ct\,
 \det\at-\lap_0-\frac{\La}3\,q_2\ct\,\cq^{-1/2}
\,.\label{PF2}
\eeq
Here we have set $R=4\La$, since we are interested in
the evaluation of the induced cosmological constant.
The quantities $q_1,q_2$ are
the roots of the second-order algebraic equation
$q^2+c_1q+c_0=0$, where
\beq
c_0=\frac{144(f'V-1+\ga)}{\ga(3-\rho)^2}\,,\hs
c_1=-\frac{24[f'V-1+\ga(3-\rho)]}{\ga(3-\rho)^2}\,.
\eeq

It is well known that the one-loop effective action in gravity, as well as in
non-abelian gauge theories, is a gauge dependent one. One can use the
gauge-fixing independent effective action (for a review,
see \cite{buch}) in order to solve gauge dependence problem. It is known
that such gauge-fixing independent action agrees with
the standard one computed in the Landau gauge $(\ga=\ii,\be=0,\rho=1)$
(but only in the one-loop approximation).
Then, we write down the expressions (\ref{PF1}) and (\ref{PF2})
also in such a particular important gauge. We get, respectively,
\beq
Z^{(1)}(\ii,0,1)&=&
 \aq\frac{\det\at-\lap_1-\La\ct}{\det\at\,-\lap_2+\frac83\,\La\ct}\:\:
 \frac{\det\at-\lap_0-2\La\ct}{\det\at\,-\lap_0+V''\ct}\cq^{1/2}\,,
 \hs\hs V(\phi)=0=V'(\phi)=0\,,
\label{PFL1}
\eeq
\beq
Z^{(1)}(\ii,0,1)&=&
 \aq\frac{\det\at-\lap_1-\La\ct}
 {\det\at\,-\lap_2-\frac{4\La(f'V-2)}{3(f'V+1)}\ct}\cq^{1/2}\:
 \aq\det\at-\lap_0-\frac43\,\La\ct\cq^{-1/2}\,,
 \hs\hs f(R)=0\,.
\label{PFL2}
\eeq
Thus, we have obtained the one-loop effective action for non-minimal modified
gravity on the de Sitter background with a constant scalar field.

\section{Vacuum energy and the induced cosmological constant}\label{ICC}

In this section we will compute the induced cosmological
constant for the last two cases,
Eqs.~(\ref{PF1})and (\ref{PF2}), discussed in the previous section.
We start with the first, Eq.~(\ref{PF1}),
which represents the one-loop contribution
to the partition function of a massive scalar field
with background solution $\phi=0$,
non-minimally coupled to gravity through the arbitrary
function $f$, which depends on gravity through the scalar curvature.
The only classical solution with constant curvature
of this model is the Minkowskian one ($\La_0=0$),
since we are dealing with the case (d) and, in fact,
when $\La=\La_0=0$ we recover the last equation in (\ref{PFos1-4}).
In this case, the one-loop effective action has a
solution with constant curvature $R=4\La$,
$\La$ being the induced cosmological constant, which is related to
the vacuum energy generated by the quantum fluctuations of the
gravitational and scalar fields.
In order to calculate this vacuum energy, we impose
the one-loop effective action $\Ga$ to have a solution
with constant curvature $R=4\La$.
Correspondingly, $\La$ has to satisfy the equation
\beq
\label{ELa}
\frac{\partial\Ga}{\partial\La}
 =\frac{\partial(I+\Ga^{(1)})}{\partial\La}=0\,,
\eeq
where $I$ is the classical action in (\ref{E0}), and $\Ga^{(1)}$
the one-loop contribution. They read
\beq
\label{Ga}
I=-\frac{96\pi^2}{\La}\,,\hs\hs
 \Ga^{(1)}=-\log Z^{(1)}_{(\ga,0,\rho)}\,.
\eeq
Of course, Eq.~(\ref{ELa}) applied to the classical
action $I$ gives rise (in our case) only to a
vanishing cosmological constant $\La_0$.

The main part of the effective action, $\Ga^{(1)}$, can be conveniently evaluated
by making use of zeta-function regularization and, specifically, by computing
the zeta-functions $\ze(s|L_n)$ related to the
differential-elliptic Laplace-like operators $L_n$ \cite{Bytsenko:1994bc,eli94}.
Using the same notations as in Ref.~\cite{Cognola:2005de}, we get
\beq
\Ga^{(1)}&=&
 Q_0\at\frac94+\frac{12}{3-\rho}\ct
 +Q_1\at\frac{25}4\ct
 -\frac12\,Q_1\at\frac{25}4-\frac{6}{\ga}\ct
 -\frac12\,Q_2\at\frac{-15}4\ct
\nn\\ &&\hs
 -\frac12\,Q_0\at\frac94+\frac{12[2(\ga-1)+\ga(1-\rho)^2]}
 {\ga(3-\rho)^2}\ct
 -\frac12\,Q_0\at\frac94-\frac{3m^2}\La\ct\,,
\label{Ga1}
\eeq
where
\beq
Q_n(\al)=\ze'\at0|L_n/\mu^2\ct\,,\hs\hs
L_n=-\lap_n-\frac\La3\,(\al-\al_n)\,,
\eeq
\beq
\ze'(0|L_n)=\lim_{s\to0}\,\frac{d}{ds}\,
\ze(s|L_n)=-\log\det L_n\,,
\eeq
and $m^2=V''$, $\al_0=9/4$, $\al_1=13/4$, $\al_2=17/4$.

 From (\ref{ELa}), we obtain the equation
for the induced cosmological constant $\La$ \cite{frad}.
In order to get it, we first observe that, even when the
$Q_n$-functions do not explicitly depend on $\La$, they
provide contributions to the induced cosmological constant
through the corresponding logarithmic term(see also Appendix in
Ref.~\cite{Cognola:2009jx}), since
\beq
\label{dQn}
\frac{d Q_n(\al)}{d\La}=
 \frac{F_\al(0)+b_0+b_1\al+\frac12\,b_2\al^2}
 {\La}+Q'_n(\al)\,\frac{d\al}{d\La}\,.
\eeq
A trivial example of this kind is the one corresponding to pure
Einstein gravity. For such case, $\Ga^{(1)}$ is given by
equation (\ref{Ga1}) without the last term on the right hand side.
Then, none of the $Q_n$-functions depends on $\al$, and we get
\beq
\label{Eins}
0=\frac{\partial\Ga}{\partial\La}
 =\frac{192\pi^2}{\La^2}-\frac{C_E}{\La}\segue
\frac1{\La_E}=\frac{C_E}{192\pi^2}\,,
\eeq
where $\La_E$ is the induced cosmological constant one obtains in pure Einsteinian gravity,
and $C_E$ a gauge-dependent constant which can be
easily computed in closed form by using (\ref{Ga1}),
(\ref{dQn}). It reads
\beq
\label{C0}
C_E=\frac{\ga_0+\ga_1\ga+\ga_2\ga^2}{180\ga^2(3-\rho)^4}\,.
\eeq
\beq
\ga_0&=&-270(259-324\,\rho+162\,\rho^2-36\,\rho^3+3\,\rho^4\,,
\nn\\
\ga_1&=&360(297-360\,\rho+176\,\rho^2-36\,\rho^3+3\,\rho^4\,,
\nn\\
\ga_2&=&-388719+537732\,\rho-282906\,\rho^2
 +66948\,\rho^3+6479\,\rho^4\,.
\eeq
A formula for $\La$ analog to the one in (\ref{Eins}) is valid also
in the presence of the scalar field when $m=0$.
In fact, in this case too the $Q_n$-functions in (\ref{Ga1}) do not depend
on $\La$, and thus we obtain
\beq
\label{m=0}
\frac1\La=\frac{C_E-\frac{61}{180}}{192\pi^2}=
\frac1{\La_E}-\frac{61}{34560\pi^2}<\frac1{\La_E}\,.
\eeq
 From the latter equation we see that the contribution due to the
coupling does not depend on the gauge parameters and it
increases the cosmological constant, as compared to
the one induced by pure Einsteinian gravity.

If $m\neq0$ the last term in (\ref{dQn}) is non-vanishing
and the equation for $\La$ is more complicated.
We assume the parameter $\al$ to be sufficiently small order that
the series can be dropped off. Then, we get
\beq
\label{mneq0}
\frac{\partial\Ga}{\partial\La}
=\frac{192\pi^2}{\La^2}-\frac{C_E+\frac{119}{180}}{\La}
 +\frac{5m^2}{4\La^2}
 -\frac{m^2}{2\La^2}\,\log\frac{\La}{3\mu^2}
+O(m^2/\La)\,,
\eeq
It needs to be noticed that, in the limit $m\to0$, the latter
equation does not lead to (\ref{m=0}) implying the presence of
a zero mode. In fact, when $m=0$, the last term in
(\ref{PF1}) has a zero-mode which has to be removed in
the computation of the corresponding zeta-function.
If in (\ref{mneq0}) we disregard the terms depending on $m$,
then the contribution due to the coupling
does not depend on the gauge parameters, as in the previous case, but
in contrast with that case it decreases the cosmological constant $\La_E$.

Let us now turn to the discussion of the second case,
Eq.~(\ref{PF2}), which in general is more complicated
 than the previous one, due to the presence of the $f$ function,
which depends on $\La$.
In this case, the one-loop contribution to the effective action
can be read off from (\ref{PF2}). It has the form
\beq
\Ga^{(1)}&=&
 Q_0\at\frac94+\frac{12}{3-\rho}\ct
 +Q_1\at\frac{25}4\ct
 -\frac12\,Q_1\at\frac{25}4-\frac{6(f'V-1)}{\ga}\ct
 -\frac12\,Q_2\at\frac{33}4-\frac{12}{f'V+1}\ct
\nn\\ &&\hs
 -\frac12\,Q_0\at\frac{25}4\ct
 -\frac12\,Q_0\at\frac94+q_1\ct
 -\frac12\,Q_0\at\frac94+q_2\ct\,.
\label{Ga2}\eeq
The computation simplifies in the case in which
$f'$ does not depend on $\La$. This means that
one considers only functions linear in the curvature.
In such case, none of the arguments of the $Q_n$-functions
in (\ref{Ga2}) depends on $\La$ and thus,
as in (\ref{m=0}), one gets
\beq
\frac{\partial\Ga}{\partial\La}
 =\frac{192\pi^2}{\La^2}-\frac{C}{\La}\segue
\frac1\La=\frac{C}{192\pi^2}\,.
\eeq
where $C=C(\ga,\rho,f'V)$ is a complicated function which, in principle,
can be computed explicitly. Here we write it down only in the Landau gauge.

Starting from (\ref{PFL2}) we have
\beq
\Ga^{(1)}=
 \frac12\,Q_1\at\frac{25}4\ct
 -\frac12\,Q_2\at\frac{33}4-\frac{12}{f'V+1}\ct
 -\frac12\,Q_0\at\frac{25}4\ct\,,
\label{Ga3}
\eeq
and choosing again $f'$ to be a constant we finally get
\beq
 \frac1\La=\frac{C}{192\pi^2}\,,\hs\hs
C=\frac{43(f'V)^2-114f'V+443}{20(f'V+1)^2}\,.
\eeq
Thus, we have explicitly found the induced cosmological constant which may be
relevant in the inflationary epoch, for several simple variants of
non-minimal modified gravity.

\section{Induced cosmological constant in the large $R$ limit}\label{LRL}

It is interesting to note that quantum corrections may indeed induce a de Sitter
universe, which generates the inflationary phase. We then consider the one-loop
effective action for a simple, specific model in the large-$R$ limit.
This means that we look for de Sitter solutions with very large curvature,
which could in fact be induced by quantum corrections.
In (\ref{Z1}), for simplicity, we set $\be=0$.

We first consider a simple toy model, quadratic in the curvature,
defined by $f(R)=\al R^2$, $\al$ being a constant.
As it follows from (\ref{dSG}), such model has only one classical solution with
constant curvature, which corresponds to Minkowski one($R_0=0$).
In the large $R=4\La$ limit, we trivially get
\beq
Z^{(1)}(\ga,0,\rho)&\sim& \aq\det\at-\lap_0+V''\ct\cq^{-1/2}\,,
\eeq
from which it follows that
\beq
\Ga=-384\pi^2\al V-\frac{24\pi^2}\La
-\frac12\,Q_0\at\frac94-\frac{3V''}{\La}\ct\,,
\eeq
\beq
\frac{\partial\Ga}{\partial\La}
=\frac{192\pi^2}{\La^2}+\frac{5V''}{4\La^2}
 -\frac{V''}{2\La^2}\,\log\frac{\La}{3\mu^2}
+O(V''/\La)\,,
\eeq
and this implies that quantum corrections induce a de Sitter solution
with large curvature
\beq
R=\frac{3}{4\mu^2}\,\exp\at\frac52+\frac{384\pi^2}{V''}\ct\,.
\eeq
The solution is stable if $V''>0$.

As a second example, we choose $f(R)=\al R\log\frac{R}{\be}$, $\al$ and $\be$ being
constants. From (\ref{dSG}) it follows that such a model has the following classical
de Sitter solutions:
\beq
\begin{array}{lll}
1): & R_0=0\,,&\hs\hs V_0\mbox{ arbitrary}\,,\\
2): & R_0=\be\,,&\hs\hs V_0=1/\al\,,\\
3): & R_0=\be\exp\at1-\frac1{\al V_0}\ct\,,&\hs\hs V'_0=0\,.
\end{array}
\eeq
The one-loop quantum corrections in the large $R$ limit for this case give
\beq
Z^{(1)}(\ga,0,\rho)&\sim&
 \aq\det\at-\lap_0+V''+\frac{[V'(\log(R/\be)-1)]^2}{V\log(R/\be)}\ct\cq^{-1/2}\,.
\eeq
Now, we obtain
\beq
\Ga=-\frac{24\pi^2}\La\at1+4\al V\log\frac{4\La}{\be}\ct
 -\frac12\,Q_0\at\frac94-\frac{3V''}{\La}
 -\frac{3[V'(\log(4\La/\be)-1)]^2}{\La V\log(4\La/\be)}\ct\,,
\eeq
\beq
\frac{\partial\Ga}{\partial\La}
&=&-\frac{96\pi^2\al V}{\La^2}+\frac{24\pi^2}{\La^2}\,\at1+4\al V\log\frac{4\La}{\be}\ct
\nn\\&&\hs
-\frac{
\aq -1.53+\at\log\frac{4\La}{\be}\ct^2\,\at6.59-3\log\frac{\La}{3\mu^2}\ct
+\log\frac{\La}{3\mu^2}+\log\frac{4\La}{\be}\,
 \at-2.53+\log\frac{\La}{3\mu^2}\ct\cq(V')^2}
 {2\La^2 V\at\log\frac{4\La}{\be}\ct^2}
\nn\\&&\hs\hs
-\frac{\at\log\frac{4\La}{\be}\ct^3
 \at-2.53+\log\frac{\La}{3\mu^2}\ct(V')^2
 +\at\log\frac{4\La}{\be}\ct^2\at-2.53+\log\frac{\La}{3\mu^2}\ct VV''
}
{2\La^2 V\at\log\frac{4\La}{\be}\ct^2}
\nn\\&&
-\frac{1.32\at\log\frac{4\La}{\be}-2.11\ct\,\at\log\frac{4\La}{\be}-1\ct\,
 \at\log\frac{4\La}{\be}+0.36\ct(V')^2+1.32\at\log\frac{4\La}{\be}\ct^2VV''}
 {2\La\log\frac{4\La}{\be}\aq\at\log\frac{4\La}{\be}-1\ct^2(V')^2+VV''\cq}\,.
\eeq
One can see (graphically) that there are values
of $R=4\La$ for which the latter expression vanishes.
In particular, for very large $R$ the last term is dominant and
$\log(4\La/\be)$ can be derived in closed form as a function of the potential and its
derivatives. In this way we get a de Sitter solution induced by quantum corrections,
which is unstable for suitable potentials and
could describe the inflationary phase of the universe.
Thus, we have demonstrated that quantum effects in non-minimal gravity may
induce a large cosmological constant which produces the de Sitter inflation.

\section{Conclusion}

In summary, we have discussed in this paper non-minimal modified gravity and its
availability for the description of the early- and late-time eras of our universe.
The one-loop effective action in the de Sitter background with a constant
background scalar
field was found. Its gauge-fixing independent version has also been obtained.
Using the one-loop effective action we derived the effective induced
cosmological constant, which may be responsible for inducing the
early-time de Sitter epoch. Moreover, it turns out that in some cases when
there is no classical de Sitter universe solution, the one-loop quantum
effects induce the de Sitter space at the quantum-corrected level.
The classical reconstruction scheme for non-minimal modified gravity was
developed and it was demonstrated that, by using such a scheme, one can present a
convenient model which unifies early-time inflation with late-time acceleration.

It would be interesting to exhibit realistic models of such unification
which turns out to be quite complicated in its explicit form. Then, the
one-loop calculation may give a way to estimate the precise role of quantum
corrections in the early-time inflation.
Unfortunately, to realize this program one has to carry out the explicit one-loop
calculation on the de Sitter background with a dynamical background
scalar, what is not the easy task. In the present work we have been able
to do
such calculation only in the case when the background scalar is constant.

>From another side, the quantum effects of non-minimal modified gravity
may be relevant also in the future for the models which develop the
classical finite-time singularity. It is to be expected that, in analogy with
the results in Refs.~\cite{abdalla, eli1} such quantum gravity effects
will drive the future
universe to the de Sitter era before the singularity occurs.. In this
respect, the one-loop calculation on the de
Sitter background can be also relevant for the future dark energy epoch.

\section*{\bf ACKNOWLEDGMENTS}
GC acknowledges the support received from
the European Science Foundation (ESF) for the activity entitled
``New Trends and Applications of the Casimir Effect''
(Exchange Grant 2262) for the period March-April 2009, during
which part of the present paper has been written. EE's research was performed
in part while on leave at Department of Physics and Astronomy, Dartmouth
College, 6127 Wilder Laboratory, Hanover, NH 03755, USA.
The work by S.N. is supported by the Global
COE Program of Nagoya University provided by the Japan Society
for the Promotion of Science (G07). The research by SDO has been supported in
part by JSPS (Japan). This paper is partly an outcome of the
collaboration program INFN (Italy) and CSIC (Spain). It has been also
supported in part by MEC (Spain), project FIS2006-02842 and grant PR2009-0314, by AGAUR
(Gene\-ra\-litat de Ca\-ta\-lu\-nya), contract 2005SGR-00790, and by RFBR,
grant 06-01-00609 (Russia).

\end{document}